\documentclass[
aps,
physrev,
reprint,
amsmath,
amssymb,
a4paper,
nofootinbib
]{revtex4-2}

\usepackage{graphicx}
\usepackage{booktabs}
\usepackage{bm}
\usepackage{wasysym}
\usepackage{mathtools}

\usepackage{hyperref}
\usepackage[noabbrev]{cleveref}

\newcommand{\e}{\text{e}}
\newcommand{\p}{\text{p}}

\begin{document}
	
\title{Sympathetic cooling of charged particles in Penning traps using electron cyclotron radiation}

\author{Jost Herkenhoff}
\email{jost.herkenhoff@mpi-hd.mpg.de}
\affiliation{Max-Planck-Institut für Kernphysik, Saupfercheckweg 1, 69117 Heidelberg, Germany}

\author{Jonathan Notter}
\affiliation{Max-Planck-Institut für Kernphysik, Saupfercheckweg 1, 69117 Heidelberg, Germany}

\author{Klaus Blaum}
\affiliation{Max-Planck-Institut für Kernphysik, Saupfercheckweg 1, 69117 Heidelberg, Germany}

\date{\today}

\begin{abstract}
	We present a new technique for cooling arbitrary charged particles in a Penning trap by utilizing self-cooled electrons stored in a separate, macroscopically distant Penning trap as the cooling medium. The electrons decay predominantly to their motional ground state by emission of cyclotron radiation, which results in extremely low temperatures in the realm of single-digit quantum numbers in the motional degrees of freedom of the sympathetically cooled particle species. This opens up an exciting new frontier of tests of fundamental physics in Penning traps. This article provides a conceptual overview as well as a quantum-mechanical description of the involved cooling dynamics.
	The first implementation of this technique is currently being realized at the dedicated ELCOTRAP experiment at the Max Planck Institute for Nuclear Physics, which introduces special features for a quick iterative technical development cycle. Its current status, first results from commissioning, and future prospects will be presented.
\end{abstract}
\maketitle

\section{Introduction}\label{sec:introduction}

Over the last decades, the field of experimental atomic, molecular and optical (AMO) physics has shown remarkable progress in controlling and cooling of individual particles, resulting in a steady trend towards ever lower temperatures.
Precision experiments involving Penning traps are no exception and became an invaluable platform for a broad physics program \cite{blaumPerspectivesTestingFundamental2020}, including stringent tests of quantum electrodynamics \cite{blaumGfactorExperimentsSimple2009} or charge-parity-time reversal symmetry \cite{gabrielseAntiprotonMassMeasurements2006, smorraBASEBaryonAntibaryon2015}, contributions to a next generation time standard using highly-charged ions \cite{kozlovHighlyChargedIons2018, kromerObservationLowLyingMetastable2023} or the search for physics beyond the Standard Model \cite{berengutProbingNewLongRange2018, wilzewskiNonlinearCalciumKing2025, gastaldoElectronCapture163Ho2017, klugeNewPromisesDetermination2007, eliseevSearchResonantEnhancement2012, doorProbingNewBosons2025}.
However, in many Penning-trap experiments the achievable precision is still limited by the particle temperature:
Leading contributions to both systematic and statistical uncertainties stem from field imperfections \cite{ketterFirstorderPerturbativeCalculation2014} or, with the relative experimental precision now routinely surpassing $10^{-10}$, relativistic effects \cite{ketterClassicalCalculationRelativistic2014}. These phase-space volume dependent effects are especially dominant for light particles, rendering lower particle temperatures highly desirable for measurements on light ions \cite{heisseHighprecisionMassSpectrometer2019, sasidharanPenningTrapMassMeasurement2023} or (anti)protons  \cite{ulmerHighprecisionComparisonAntiprotontoproton2015,smorraPartsperbillionMeasurementAntiproton2017}.
In addition, the particle temperature directly impacts the state detection of nuclear \cite{schneiderNovelPenningTrapDesign2019} or electronic \cite{vogelAnomalousMagneticMoment2009,sturmALPHATRAPExperiment2019} magnetic moments in experiments utilizing the continuous Stern-Gerlach effect (CSGE)  \cite{vandyckElectronMagneticMoment1981}, and has been the primary limitation in a recently demonstrated measurement combining the CSGE with laser spectroscopy \cite{eglApplicationContinuousSternGerlach2019} due to Doppler broadening.

In response to the apparent demand for lower particle temperatures, we propose a new technique to cool arbitrary charged particles down to single-digit quantum numbers in the motional degrees of freedom by using the natural cyclotron radiation of electrons as a cooling mechanism:
The cyclotron eigenmotion of an electron in a strong magnetic field sufficiently couples to the electromagnetic field and therefore eventually thermalizes with the black-body photons of its environment, which is typically cooled to $\sim 4\,\text{K}$ in cryogenic Penning-trap experiments. The average motional quantum number associated with this motion then approaches $\bar n \approx 0.1$, indicating that electrons are generally very appealing to use as a cooling medium.
In order to make this cooling process available to arbitrary particles-of-interest, the electrons' cold cyclotron motion first has to be coupled to an orthogonal eigenmotion of the electron, whose frequency can be precisely matched to a range of particles-of-interest. This coupling is accomplished by driving a sideband transition in the millimeter-wave regime, which demands a carefully optimized Penning-trap design.
The particle-of-interest is then stored in a separate, macroscopically distant Penning trap and is coupled to the well-cooled electron using image charge interaction. This type of sympathetic coupling was proposed by Heinzen and Wineland already in 1990 \cite{heinzenQuantumlimitedCoolingDetection1990}
and has recently seen experimental progress through successful coupling of ions stored in two separate Penning traps \cite{bohmanSympatheticCoolingTrapped2021, tuTankCircuitAssistedCoupling2021}, paving the way for our proposed cooling technique.

The development and first implementation of this technique is carried out at the dedicated ELCOTRAP  (Electron Cooling Trap) experiment at the Max Planck Institute for Nuclear Physics in Heidelberg, Germany. The experimental architecture was specifically optimized to support an iterative development process, which allows different aspects of the proposed cooling technique to be tested in successive experimental phases. While being primarily designed for the present use case, the flexibility of this experiment will serve as a useful tool for future technical developments in the field of Penning-trap experiments.

In this manuscript, we first give a conceptual overview of the proposed cooling technique, which is then followed by a detailed theoretical description. Finally, the experimental apparatus is described and first results from commissioning are presented.

\newpage
\section{Experimental foundation}
\label{sec:overview}

\begin{figure*}
	\includegraphics[width=\textwidth]{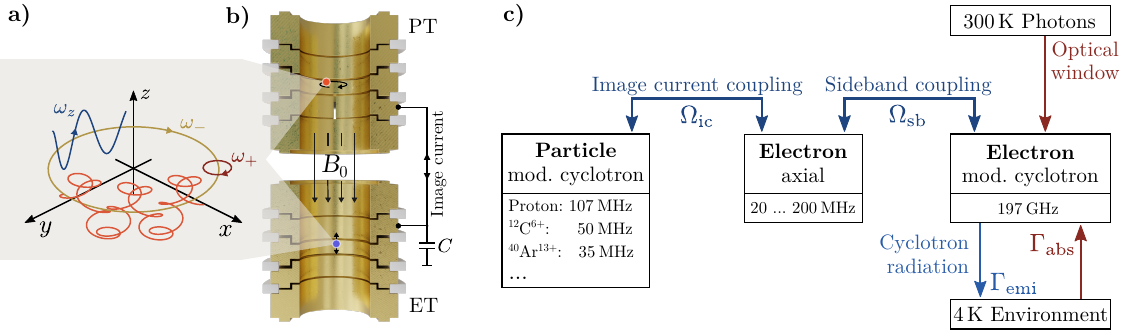}
	\caption{Inset \textbf{a)} depicts the full trajectory of a charged particle stored in a Penning trap in bright red, with its decomposed individual eigenmotions being drawn separately in different colors. Figure \textbf{b)} depicts the coupling of particles stored in separate Penning traps, in which a radially split electrode of the particle trap (PT) picks up the modified cyclotron motion of the particle-of-interest in the form of an image charge and couples it to the axial motion of the electron via electrical connection to an axially offset electrode of the electron trap (ET). Figure \textbf{c)} shows a block diagram of the full cooling chain between the three involved eigenmotions. For each motion, the eigenfrequencies are listed for a magnetic field of $7\,\text{T}$ and in case of the electron's axial motion for typical Penning-trap parameters and varying electrostatic potential depths. The red/blue arrows indicate the cooling/heating channels, while the direction of the arrows encode the primary energy flow during cooling.}
	\label{fig:coupling_overview}
\end{figure*}

Penning traps utilize a homogeneous magnetic field 
\begin{equation}
\label{eq:magnetic_field}
\bm{B} = B_0 \bm{e}_z\,,
\end{equation}
without loss of generality defined to point along the z-axis, to confine a particle with charge $q$ and mass $m$ onto a circular orbit with angular frequency $\omega_c = q B_0/m$ perpendicular to $\bm{e}_z$. Superimposing an electrostatic potential 
\begin{equation}
\label{eq:quadrupole_potential}
\phi = -\frac{1}{2}C_2 V_0 \left(x^2 + y^2 - 2z^2\right)
\end{equation}
generated by a stack of electrodes gives rise to confinement of the particle in all three spatial dimensions. The constant $C_2$ is defined by the geometry of electrodes and $V_0$ is the amplitude of the applied voltage.
The particle follows the trajectory shown in figure \hyperref[fig:coupling_overview]{\ref{fig:coupling_overview}a}, which is composed of a harmonic \emph{axial} oscillation with frequency
\begin{equation}
\label{eq:axial_frequency}
\omega_z = \sqrt{\frac{2 q C_2 V_0}{m}}
\end{equation}
along the magnetic field lines as well as two independent oscillations perpendicular to the magnetic field with frequencies
\begin{equation}
\label{eq:radial_frequencies}
\omega_{\pm} = \frac{1}{2}(\omega_c \pm \sqrt{\omega_c^2 - 2\omega_z^2})\,,
\end{equation}
called the \emph{modified cyclotron} ($\omega_+$) and \emph{magnetron} ($\omega_-$) motions. Typical eigenfrequencies for different particle species are listed in \cref{tab:eigenfrequencies} for comparison.

\begin{table}
	\caption{Comparison of eigenfrequencies for different particle species. The listed modified cyclotron frequencies ($\omega_+$) are calculated for a magnetic field strength of $7\,\text{T}$, while the axial frequencies ($\omega_z$) are taken from the listed experimental references. The axial frequency of an electron can be varied over a wide range by adjusting the trap depth $V_0$, which allows it to be matched to the modified cyclotron frequencies of various particles-of-interest, as indicated by the underlined entries.}
	\label{tab:eigenfrequencies}
	\setlength{\tabcolsep}{5pt}
	\begin{tabular}{lrrr}
		\toprule
		& $\omega_+/2\pi$ & $\omega_- / 2\pi$ & $\omega_z / 2\pi$ \\
		\midrule
		Electron             & $197\,\text{GHz}$         & $2 \text{ to } 100\,\text{kHz}$             & $\underline{20 \text{ to } 200\,\text{MHz}}$ \\
		(Anti)Proton         & $\underline{107\,\text{MHz}}$         & $2\,\text{kHz}$             & $675\,\text{kHz}$  \cite{smorraPartsperbillionMeasurementAntiproton2017} \\
		$^{4}\text{He}^{2+}$ & $\underline{54\,\text{MHz}}$          & $2\,\text{kHz}$             & $468\,\text{kHz}$  \cite{sasidharanPenningTrapMassMeasurement2023} \\
		$^{12}\text{C}^{6+}$ & $\underline{54\,\text{MHz}}$          & $2\,\text{kHz}$             & $525\,\text{kHz}$  \cite{heisseHighprecisionMassSpectrometer2019} \\
		$^{40}\text{Ar}^{13+}$ & $\underline{35\,\text{MHz}}$          & $6\,\text{kHz}$             & $650\,\text{kHz}$  \cite{eglApplicationContinuousSternGerlach2019} \\
		$^{118}\text{Sn}^{49+}$ & $\underline{45\,\text{MHz}}$          & $5\,\text{kHz}$             & $650\,\text{kHz}$  \cite{morgnerStringentTestQED2023} \\
		\bottomrule
	\end{tabular}
\end{table}

It is well known that an accelerated non-relativistic charged particle emits electromagnetic radiation into free space at a rate proportional to the acceleration-squared  \cite{larmorLXIIITheoryMagnetic1897}, indicating that it is mostly relevant for particles oscillating at high frequencies. In fact, the only particle that shows significant radiation in a Penning trap is an electron, whose very fast modified cyclotron motion gives rise to the so called \emph{cyclotron radiation}, which is described in detail in  \cref{sec:theory:cyclotron_radiation}. This process causes its modified cyclotron motion to decay, counteracted only by absorption of photons from its environment. Taking care to minimize any excessive photon influx, this ideally reduces to the black-body background and thus leads to thermalization to the temperature of its environment.
By cooling the Penning trap to $4\,\text{K}$, the average motional quantum number associated with the electron's modified cyclotron motion in thermal equilibrium becomes 
\begin{equation}
\label{eq:average_quantum_number_electron}
\bar n_{+,\e} = \frac{1}{e^{\hbar \omega_{+,\e} / k_\text{B} T} - 1} \approx 0.1\,,
\end{equation}
where the subscripts in $n_{+,\e}$ and $\omega_{+,\e}$ were introduced to distinguish them from the same set of variables associated with the particle-of-interest, which will be subscripted with $\p$.
Unfortunately, this well-cooled motion is not readily available to sympathetically cool arbitrary particles due to an $\mathcal{O}(10^3)$ mismatch between the electron's modified cyclotron frequency and even the highest eigenfrequency of possible particles-of-interest, as can be seen in \cref{tab:eigenfrequencies}.
To bridge this large frequency gap, the electron's cyclotron motion is first coupled to its axial motion by driving its motional sideband \cite{cornellModeCouplingPenning1990}. This results in a periodic energy exchange with angular frequency $\Omega_\text{sb}$ between the modes, analogous to the well-known Rabi-flopping of two level systems. During this process, the electron's modified cyclotron motion continues to dissipate energy through cyclotron radiation, which leads to an effective cooling of the axial motion until $\bar n_{z,\e} = \bar n_{+,\e} \approx 0.1$.
Sideband coupling of the modified cyclotron to the axial motion is a routinely employed technique, which is commonly implemented by applying a radio-frequency drive to a radially split electrode of the Penning trap, which produces an electric field with components proportional to $x \bm{e}_z$ or $z \bm{e}_x$ at the center of the trap. 
As the particle moves through the spatially varying field, it experiences a modulation of the applied radio-frequency drive that leads to resonant transitions between the two motions when driving the sideband $\omega_\text{sb} = \omega_{+} - \omega_{z}$.
For the proposed cooling technique, the electron's modified cyclotron motion is coupled to its axial motion using a sideband drive at $\omega_\text{sb} /2\pi \approx 196\,\text{GHz}$. Due to this very high frequency in the millimeter-wave regime, the implementation using a split electrode is no longer feasible as impedance matching to the electrode and signal transition using conventional cables become almost impossible.
Instead, the required field gradient is produced by injecting a traveling wave into the Penning trap, which effectively acts as a circular waveguide and therefore is very effective at propagating millimeter-waves if designed appropriately.

The axial frequency of electrons can be freely tuned from $\omega_{z,\e}/2\pi \approx 20 \text{ to } 200\,\text{MHz}$ for typical Penning-trap parameters by varying the applied voltage $V_0$ to the trap electrodes. This happens to cover the range of modified cyclotron frequencies of many possible particle species, including highly-charged ions, light ions or protons and antiprotons, as highlighted in \cref{tab:eigenfrequencies}. This allows the electron's well-cooled axial motion to be precisely matched to the modified cyclotron motion of a particle-of-interest, which opens up the possibility for sympathetic cooling. For this, the particle-of-interest is stored in a separate Penning trap adjacent to the electron's trap. The initially thermally excited modified cyclotron motion of the particle-of-interest induces image charges within its trap's electrodes, which can be picked up by radially splitting one of the electrodes in half. Electrically connecting one of these halves to an axially offset electrode of the electron trap using a simple wire, as shown in \hyperref[fig:coupling_overview]{\ref{fig:coupling_overview}b}, results in an effective coupling to the electron's axial motion. This image-charge induced interaction leads to periodic energy exchange with frequency $\Omega_\text{ic}$ between the particles modified cyclotron motion and the electrons axial motion. This finally results in a thermalization of the particle's modified cyclotron motion to the same low average quantum number of $\bar n_{+,\p} \approx 0.1$ prescribed by the electron's modified cyclotron motion. A block diagram of the full chain of coupled eigenmotions is depicted in figure \hyperref[fig:coupling_overview]{\ref{fig:coupling_overview}c}.
Evaluating the temperature associated with $\bar n_{+,\p}$ through \cref{eq:average_quantum_number_electron} reveals a range of $T_{+,\p} \approx 0.7 \text{ to } 2.2\,\text{mK}$ in the modified cyclotron motion for possible particles-of-interest listed in \cref{tab:eigenfrequencies}.

This result provides a promising insight into the proposed cooling technique and highlight its capabilities to cool an almost arbitrary range of particles-of-interest to temperatures that could not be reached in Penning traps with the exception of direct laser cooling of a few selected ion species.
However, the previous simplified description neglects any parasitic heating effects which inevitably will lead to a degradation of cooling performance. It is therefore of great importance to identify and minimize potential heating channels, which might stem from influx of room-temperature black-body radiation or electrical noise within the trap electronics.
In order to reduce the impact of any residual heating effects, the cooling chain must be optimized for highest possible cooling rates.

Another important practical consideration is the time required to reach the desired temperature. In most measurement schemes the particle needs to be cooled for each recorded data point, making rapid cooling desirable to acquire sufficient statistics. While the cooling time of conventional techniques like resistive cooling, buffer-gas cooling or direct laser cooling can be analytically quantified as an exponential decay of the particle's energy, the proposed cooling technique does not lend itself to such a straightforward analytical treatment.
Yet, to quantify the cooling timescale of the proposed technique, the next section provides analytical descriptions of the individual coupling mechanisms, which are then combined through numerical simulations to explore the full cooling dynamics.

\section{Theoretical description}
\label{sec:theory}

The commonly used classical treatment of Penning traps is often justified by the fact that the particles' average motional quantum numbers $\bar n = (\exp(\hbar \omega / k_\text{B}T)-1)^{-1}$ exceed $10^3$ at currently routinely achieved particle temperatures. However, as the proposed cooling technique involves dynamics with $\bar n < 1$, we derive the dynamics of the proposed cooling scheme in a quantized formalism.
Beyond that, the spontaneous emission of cyclotron radiation is inherently dependent on vacuum fluctuations and thus relies on a quantized description of the electromagnetic field.

The description of a particle's motion in a Penning trap is based on the minimal-coupling Hamiltonian \cite{marlano.scullyQuantumOptics1997}
\begin{equation}
\label{eq:theory:minimalCouplingHamiltonian}
\hat{H}_0 = \frac{1}{2m} \left(\hat{\bm{p}} - q \hat{\bm{A}}_0(\hat{\bm{r}})\right)^2 + q\hat{\phi}(\hat{\bm{r}})
\end{equation}
which neglects the spin-degree of freedom as it is not relevant in the present case. The magnetic vector potential operator $\hat{\bm{A}}_0 = -B_0/2 (\hat{y} \bm{e}_x - \hat{x} \bm{e}_y)$ reproduces the magnetic field from \cref{eq:magnetic_field} and is chosen to satisfy the Coulomb gauge condition \cite{griffithsIntroductionElectrodynamics2017} $\nabla \cdot \hat{\bm{A}}_0 = 0$ and rotational symmetry around $\bm{e}_z$. The potential operator $\hat{\phi}$ is given by \cref{eq:quadrupole_potential} with position variables replaced by their operator counterparts. The coupling between the $\hat{x}/\hat{y}$ degrees-of-freedom introduced by the magnetic field is eliminated using a canonical transformation, resulting in three independent harmonic oscillators in the transformed phase space. This Hamiltonian is then readily diagonalized using three sets of bosonic creation and annihilation operators $\hat{a}^\dagger_i$ and $\hat{a}_i$ acting on their respective Fock space $\mathcal{F}_i$, with $i \in \{+, -, z\}$ designating the modified cyclotron, magnetron and axial degrees-of-freedom known from classical Penning-trap theory.
The operators obey the commutation relations $[\hat{a}_i, \hat{a}^\dagger_j] = \delta_{ij}$. The final Hamiltonian then takes the simple form \cite{brownGeoniumTheoryPhysics1986}
\begin{equation}
\begin{split}
	\label{eq:theory:hamiltonianCreationAnnihilation}
	\frac{\hat{H}_0}{\hbar} = \omega_+ \left(\hat{a}_+^\dagger\hat{a}_+ + \frac{1}{2}\right) &+ \omega_z \left(\hat{a}_z^\dagger\hat{a}_z + \frac{1}{2}\right) \\
	&- \omega_- \left(\hat{a}_-^\dagger\hat{a}_- + \frac{1}{2}\right)\,,
\end{split}
\end{equation}
with eigenfrequencies of the individual modes given by \cref{eq:axial_frequency,eq:radial_frequencies}.

\subsection{Cyclotron radiation in a cylindrical waveguide}
\label{sec:theory:cyclotron_radiation}

The total power radiated by an accelerated electron into free space is classically calculated using the Larmor formula \cite{larmorLXIIITheoryMagnetic1897}.
However, a Penning trap differs from free space in the sense that its conducting electrodes impose boundary conditions on the radiation field, which alters the electron's radiation dynamics. In the following, the rate of radiation in a Penning trap is derived from first principles by calculating the transition rates between individual excitations of the electron's modified cyclotron motion and photons in the radiation field. The latter is modeled by a decomposition into modes of a cylindrical waveguide, which closely resembles a cylindrical, open-endcap Penning trap.

The system is described by
$\hat{H} = \hat{H}_0 + \hat{H}_f + \hat{H}'$, with $\hat{H}_0$ representing the unperturbed dynamics of the electron in the Penning trap defined in \cref{eq:theory:hamiltonianCreationAnnihilation}, $\hat{H}_f$ being the free Hamiltonian of the radiation field and $\hat{H}'$ representing the interaction between electron and field.
The latter is derived by starting with the minimal coupling Hamiltonian defined in \cref{eq:theory:minimalCouplingHamiltonian} and substituting $\hat{\bm{A}}_0 \rightarrow \hat{\bm{A}}_0 + \hat{\bm{A}}'$, which introduces the magnetic vector potential $\hat{\bm{A}}'$ of the radiation field. Separating out the unperturbed dynamics $\hat{H}_0$ results in the interaction Hamiltonian
\begin{equation}
\label{eq:interaction hamiltonian}
\hat{H}' = \frac{q^2}{2m} \left\{\hat{\bm{A}}_0, \hat{\bm{A}}'\right\} + \frac{q}{2m}\left\{\hat{\bm{p}}, \hat{\bm{A}}'\right\} + \frac{q^2}{2m} \hat{\bm{A}}'^2\,,
\end{equation}
where $\{\cdot, \cdot\}$ denotes the anti-commutator. The diamagnetic term proportional to $\hat{\bm{A}}'^2$ is neglected in the following as the field's amplitude is very small.
The radiation field is quantized by expressing its magnetic vector potential as a field operator in mode decomposition as \cite{orszagQuantumOptics2000}
\begin{equation}
\label{eq:magnetic_vector_potential}
\hat{\bm{A}}' (\hat{\bm{r}}, t) = \sum_{s\sigma} \sqrt{\frac{\hbar}{2\omega_{s\sigma} \epsilon_0}} \left[ \hat{a}_{s\sigma}(t) \bm{u}_{s\sigma}(\hat{\bm{r}}) + \text{H.c.}\right]\,,
\end{equation}
where H.c.\ represents the Hermitian conjugate and $\bm{u}_{s\sigma}(\hat{\bm{r}})$ are vector valued functions describing the field profiles of modes indexed by $s\sigma$ with frequency $\omega_{s\sigma}$.
The operators $\hat{a}_{s\sigma}(t) = \hat{a}_{s\sigma}(0) e^{-i\omega_{s\sigma} t}$ and its hermitian conjugate annihilate/create a photon in the corresponding mode, where the time dependence of the free evolution with respect to $\hat{H}_f = \sum_{s\sigma} \hbar \omega_{s\sigma} (\hat{a}_{s\sigma}\hat{a}_{s\sigma}^\dagger + \frac{1}{2})$ was explicitly absorbed into the operators.

The mode functions $\bm{u}_{s\sigma}(\hat{\bm{r}})$ are found from solutions of the time-harmonic Maxwell equations within an infinitely long, perfectly conducting cylinder with radius $\rho_0$ along $\bm{e}_z$ to approximate a cylindrical Penning trap.
\Citet{kakazuFieldQuantizationSpontaneous1996} published a procedure for quantizing the field of a cylindrical cavity, whose closed axial boundaries lead to standing wave solutions that can be interpreted as a superposition of two counter-propagating traveling waves in an infinitely long cylinder. Therefore, replacing the discrete standing-wave dependence along the longitudinal axis by a traveling-wave factor $\exp (-i  k_z z)$ while retaining the transverse mode structure yields the solutions
\begin{equation}
	\label{eq:helmholtz_solution}
	\psi_{s\sigma} = c_{s\sigma} J_m\left(\frac{\chi_{s\sigma} \rho}{\rho_0}\right) e^{i (\phi m - k_z z)}
\end{equation}
to the scalar Helmholtz equation for an infinitely long cylinder in cylindrical coordinates $\{\rho, \phi, z\}$.
Two orthogonal sets of mode functions $\bm{u}_{s\sigma}$, corresponding to modes that are transverse-magnetic (TM; $\sigma = 1$) or transverse-electric (TE; $\sigma = 2$) with respect to the wave propagation along $\bm{e}_z$, are then constructed from the scalar solution via
\begin{subequations}
	\label{eq:vector_mode_function_relation}
	\begin{align}
		\bm{u}_{s1} &= \nabla \times \nabla \times \psi_{s1} \bm{e}_z \\
		\bm{u}_{s2} &= \nabla \times \frac{\partial}{\partial t} \psi_{s2} \bm{e}_z\,.
	\end{align}
\end{subequations}
The solutions are labeled by the composite mode index $s = (m, n, k_z)$, where $m \in \{0, \pm 1, \pm 2, \dots\}$ and $n \in \{1, 2, 3, \dots\}$ describe the azimuthal and radial mode indices, respectively. The longitudinal wave number  $k_z$ obeys the dispersion relation
\begin{equation}
	\label{eq:dispersion_relation}
	k_z = \pm \frac{1}{c}\sqrt{\omega_{s\sigma}^2 - \omega_{0,s\sigma}^2}\,,
\end{equation}
indicating that the field of a given mode becomes evanescent below its cutoff frequency $\omega_{0,s\sigma} = \chi_{s\sigma} c / \rho_0$.
The function $J_m$ denotes the $m$-th order Bessel function of first kind, $\chi_{s1} \coloneqq \chi_{mn1}$ is the $n$-th zero point of $J_m$ and $\chi_{s2} \coloneqq \chi_{mn2}$ is the $n$-th zero point of the derivative of $J_m$.
The constants
\begin{subequations}
\label{eq:normalization_constants}
\begin{align}
c_{s1} &= \frac{1}{\sqrt{V}} \frac{\rho_0 \, c}{\omega_{s1} \chi_{s1} |J_{m+1}(\chi_{s1})|}\,, \\
c_{s2} &= \frac{1}{\sqrt{V}}\frac{\rho_0}{\omega_{s2} \chi_{s2} \sqrt{J_m^2(\chi_{s2}) - J_{m+1}^2(\chi_{s2})}}
\end{align}
\end{subequations}
are chosen to normalize the free field Hamiltonian to single photon energy within an arbitrary quantization volume $V = \pi \rho_0^2 L$ of a cylindrical section of length $L$. The quantization volume and length carry no physical significance and naturally drop out in the final calculation of the transition rates.
The normalization constants differ from those in \citet{kakazuFieldQuantizationSpontaneous1996} by a factor of $\sqrt{2}$ due to the change from standing-wave to traveling-wave modes.

Substituting \cref{eq:helmholtz_solution} into \cref{eq:vector_mode_function_relation}, the mode functions can be expanded to
\begin{widetext}
	\begin{subequations}
		\label{eq:waveguide_mode_functions}
		\begin{equation}
		\bm{u}_{s1} = c_{s1} \left[ -i \frac{\chi_{s1} k_z }{\rho_0} J'_m \left( \frac{\chi_{s1} \rho}{\rho_0}\right)  \, \hat{\bm{\rho}} + \frac{m k_z }{\rho} J_m \left( \frac{\chi_{s1} \rho}{\rho_0}\right) \, \hat{\bm{\phi}}	- \frac{\chi_{s1}^2}{\rho_0^2} J_m\left(\frac{\chi_{s1} \rho}{\rho_0}\right)  \, \hat{\bm{z}} \right] e^{i(\phi m - k_z z)}\,,
		\end{equation}
		\begin{equation}
		\bm{u}_{s2} = - \omega_{s2} c_{s2} \left[ \frac{m}{\rho} J_m\left( \frac{\chi_{s2} \rho}{\rho_0} \right) \,\hat{\bm{\rho}} + i \frac{\chi_{s2}}{\rho_0} J'_m\left(\frac{\chi_{s2} \rho}{\rho_0}\right) \, \hat{\bm{\phi}} \right] e^{i(\phi m - k_z z)}\,.
		\end{equation}
	\end{subequations}
\end{widetext}
The small spatial extent of the electron's wave function compared to the fields radial length scale $\rho_0/\chi_{s\sigma}$ motivates an expansion of $\bm{u}_{s\sigma}$ around $\rho = 0$ to zeroth order, analogous to the commonly employed dipole approximation.
Evaluating $\bm{u}_{s\sigma}$ in this limit reveals vanishing radial field components for modes with $m \neq \pm 1$, indicating that these modes do not couple to the radial motion of an electron centered in the trap.
Only considering the relevant $m = \pm 1$ modes and transforming to Cartesian coordinates at $\rho \rightarrow 0$ yields

\begin{subequations}
\label{eq:waveguide_mode_functions_center}
\begin{align}
	\tilde{\bm{u}}_{s1} \Big \vert_{m = \pm 1} &= \frac{c_{s1} \chi_{s1} k_z}{2 \rho_0} (\pm i\bm{e}_x + \bm{e}_y) e^{-ik_z z}\,,\\
	\tilde{\bm{u}}_{s2} \Big \vert_{m = \pm 1} &= \frac{c_{s2} \chi_{s2} k_z \omega_{s2}}{2 \rho_0} (\pm i \bm{e}_y -\bm{e}_x) e^{-ik_z z}\,.
\end{align}
\end{subequations}

Assuming that the field modes are coupled to a thermal reservoir and their correlations decay much faster compared to the timescale of the electron-field interaction justifies a Markovian approximation of the system \cite{breuerTheoryOpenQuantum2007}.
The transition rate between excitations of the electron's modified cyclotron motion and photons in a given waveguide mode can then be calculated using Fermi's golden rule  \cite{fermiNuclearPhysicsCourse1950}
\begin{equation}
\label{eq:golden_rule}
\Gamma_{s\sigma} = \frac{2\pi}{\hbar} \vert \langle f \vert \hat{H}' \vert i \rangle \vert^2 \rho_{s\sigma}(E_f)\,,
\end{equation}
where $|i\rangle$ and $|f\rangle$ are the initial and final states, respectively, and $\rho_{s\sigma}(E_f)$ is the density-of-states (DOS) in the continuous energy spectrum of the final state. The DOS is governed by the continuous distribution of the waveguide modes in the longitudinal wave number $k_z$.
For a single waveguide-mode, the number of states within $k_z$ to $k_z + \text{d}k_z$ is $\text{d}N(k_z) = L/2\pi\,\text{d}k$, where $L$ is an arbitrary quantization length. Using the dispersion relation in \cref{eq:dispersion_relation}, the DOS with respect to energy becomes
\begin{equation}
	\label{eq:density_of_states}
	\rho_{s\sigma}(\omega_{s\sigma}) = \frac{\text{d}N_{s\sigma}}{\text{d}E_{s\sigma}} = \frac{\omega_{s\sigma} L }{\pi \hbar c \sqrt{\omega_{s\sigma}^2 - \omega_{0, s\sigma}^2}}\,.
\end{equation}

Evaluating the relevant single-photon transition matrix elements from initial state $|n_+, n_{s\sigma}\rangle$ to final state $|n_+ - 1, n_{s\sigma} + 1 \rangle$ (emission) or $|n_+ + 1, n_{s\sigma} - 1 \rangle$ (absorption) reveals that only one of the $m = \pm 1$ modes contribute to the transition, depending on the direction of the magnetic field. This is because the sign of $m$ determines the polarization-handedness of the photon, which must be co-rotating with the modified cyclotron motion in order to conserve angular momentum.
Expanding the matrix elements in \cref{eq:golden_rule} using the previously defined initial and final states and inserting \cref{eq:density_of_states} yields the rates of emission/absorption of a cyclotron-photon into/from a single waveguide mode in the form of
\begin{subequations}
	\label{eq:cyclotron_transition_rates}
	\begin{align}
		\Gamma_{\text{emi}, s\sigma} = \gamma_{s\sigma} n_+ (n_{s\sigma} + 1)\,,
		\\
		\Gamma_{\text{abs}, s\sigma} = \gamma_{s\sigma} n_{s\sigma} (n_+ + 1)\,,
	\end{align}
\end{subequations}
respectively, with rate constants $\gamma$ given by
\begin{subequations}
	\label{eq:theory:rate_constants}
	\begin{align}
		\gamma_{s1} = \frac{e^2}{2\pi \epsilon_0 \rho_0^2 m_e c} &\frac{\sqrt{\omega_+^2 - \omega_{0, s1}^2}}{\omega_+ - \omega_-} \, \frac{1}{J_0^2 (\chi_{s1})} \,, \\
		\begin{split}
			\gamma_{s2} = \frac{e^2}{2\pi \epsilon_0 \rho_0^2 m_e c} &\frac{\omega_+^2}{(\omega_+ - \omega_-)\sqrt{\omega_+^2 - \omega_{0, s2}^2}} \\
			& \cdot\frac{1}{J_1^2 (\chi_{s2}) - J_0^2 (\chi_{s2})}\,.
		\end{split}
	\end{align}
\end{subequations}
In the vicinity of the mode cutoff frequencies, the radiation rate into the $\sigma = 2$ (TE) modes diverges, causing a breakdown of the Markovian approximation as the coupling can no longer be considered weak. While an accurate description of these regimes would require a non-perturbative treatment and the inclusion of waveguide wall losses, the present formalism remains well suited in between the singularities.
Below the cutoff frequencies, the modes turn evanescent and \cref{eq:theory:rate_constants} becomes imaginary, signaling a complete suppression of spontaneous emission.

\begin{figure}
	\centering
	\includegraphics[width=\columnwidth]{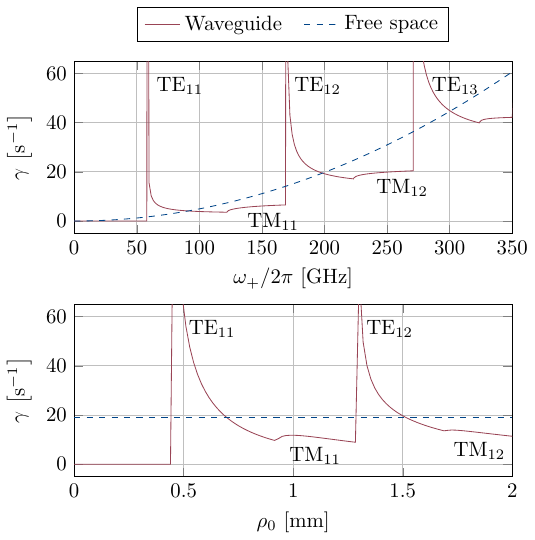}
	\caption{Rate of single-photon transitions between excitations of an electron's modified cyclotron motion and photons in a circular waveguide vs. free space. The rate is calculated by summing over the individual modes $\gamma = \sum_{s\sigma} \gamma_{s\sigma}$, while the distinct contributions from individual modes have been labeled using the convention $\text{TM}_{mn}$ for $\sigma = 1$ modes and $\text{TE}_{mn}$ for $\sigma = 2$ modes. The upper graph shows the frequency dependence for a fixed waveguide radius $\rho_0 = 1.5\,\text{mm}$, while the lower graph shows the rate for a constant frequency $\omega_+/2\pi = 197\,\text{GHz}$ and varying waveguide radius. In practice, the singularities occurring at the mode's cutoff frequencies $\omega_{0, s\sigma}$ would be ``smeared out'' by losses in the waveguide walls.}
	\label{fig:cyclotron_radiation}
\end{figure}

The total spontaneous emission rate $\gamma = \sum_{s\sigma} \gamma_{s\sigma}$ is obtained by summing over the individual mode contributions and is plotted in \cref{fig:cyclotron_radiation}. 
These results are compared to the rate of cyclotron radiation into free space by repeating the previous calculation for plane-wave mode functions defined by $\bm{u}_{\bm{k}\alpha} = \bm{\epsilon}_{\bm{k}\alpha} e^{i \bm{k} \cdot \bm{r}} / \sqrt{V}$, where $\bm{k} \in \mathbb{R}^3$ is the wave vector of the emitted photon. The polarization vector $\bm{\epsilon}_{\bm{k}\alpha}$ satisfies $|\bm{\epsilon}_{\bm{k}\alpha}| = 1$ and $\langle \bm{\epsilon}_{\bm{k}\alpha}, \bm{k}\rangle = 0$ for the two orthogonal linear polarizations denoted by $\alpha \in \{1, 2\}$.
The free-space transition rate into an infinitesimal solid angle $\text{d}\Omega$ is then calculated using \cref{eq:golden_rule} and the free-space density of states
\begin{equation}
\rho(E_f) = g_\alpha\frac{\omega_+^2 V}{\hbar (2\pi c)^3} \, \text{d}\Omega\,,
\end{equation}
where $g_\alpha=2$ accounts for the two possible linear polarizations. Integrating out the $\bm{k}$-dependence of the matrix element over the total solid angle yields the total radiation rate
\begin{equation}
	\gamma_\text{free} = \frac{\omega_+^3 q_e^2}{3 \pi \epsilon_0 c^3 m_e (\omega_+ - \omega_-)}\,.
\end{equation}

The results are plotted in \cref{fig:cyclotron_radiation} for comparison, where it can be seen that the rate of transition into the sum of waveguide modes approximately follows the general quadratic frequency behavior of transitions into free space. However, the waveguide's variation of local density of states leads to frequencies where the transition probability is either greatly enhanced or suppressed. This effect is most pronounced below the lowest cutoff frequency, where transitions are completely inhibited.  Since the cutoff-frequencies can be freely tuned by varying the waveguide radius $\rho_0$, the radiation rate might be ``engineered'' within bounds.

As the electron's modified cyclotron motion eventually reaches thermal equilibrium with the field mode, detailed balance dictates that $\bar n_+ = \bar n_{s\sigma}$.
The average photon number follows Bose-Einstein statistics
\begin{equation}
	\bar n_{s\sigma} = \frac{1}{e^{\hbar \omega_{s\sigma} / k_\text{B} T} - 1}\,,
\end{equation}
which yields $\bar n_{s\sigma} = 0.1$ assuming the field is fully thermalized to its cryogenic environment at $T \approx 4\,\text{K}$.
However, the requirement to inject a sideband drive necessitates the trap to be coupled to a millimeter-wave source positioned at room-temperature, which raises the field's average photon number due to injected Johnson-Nyquist and phase-noise. As will be presented in \cref{sec:phase_2}, a combination of room-temperature and cryogenic attenuators limits the influx of photons and leads to an effective field temperature of $T_\text{eff} \approx 8\,\text{K}$, corresponding to an average photon number of $\bar n_{s\sigma}=0.4$.

\subsection{Sideband coupling}
\label{sec:theory:sideband_coupling}

The theoretical description of sideband coupling follows a semi-classical approach using a quantized description of the particle motion while treating the externally applied sideband drive in its classical limit. The Hamiltonian is given by $\hat{H} = \hat{H}_0 + \hat{H}'$, where $\hat{H}_0$ is defined in \cref{eq:theory:hamiltonianCreationAnnihilation} and the interaction Hamiltonian $\hat{H}'$ takes the form given in \cref{eq:interaction hamiltonian}.
However, the interaction field operator $\bm{A}'$ defined in \cref{eq:magnetic_vector_potential} is now expressed as a classical field by substituting $\hat{a}_{s\sigma}(t) \rightarrow a_{s\sigma} e^{-i\omega_{s\sigma}t}$, where $a_{s\sigma}$ is a scalar amplitude that only depends on the power that is injected into the mode.
The mode functions $\bm{u}_{s\sigma}$ are given by \cref{eq:waveguide_mode_functions}, where we are again only interested in the non-vanishing $m=\pm 1$ modes in the limit $\rho \rightarrow 0$ and expand around $z=0$ up to first order. The zeroth-order term leads to dipolar excitation operators, while the first order yields the desired coupling operators. The sideband drive frequency is set to $\omega_{s\sigma} = \omega_+ - \omega_z + \delta$, with $\delta$ being a detuning from the ideal sideband frequency.

\begin{figure}
	\centering
	\includegraphics[width=\columnwidth]{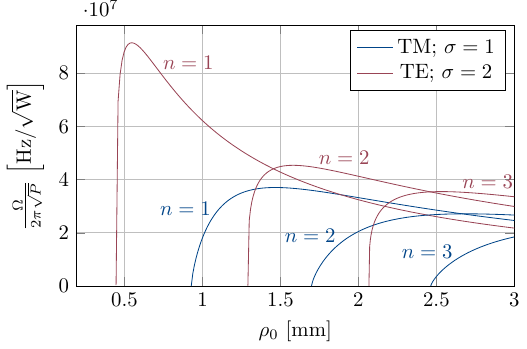}
	\caption{Coupling rate between an electron's modified cyclotron motion at $\omega_+ / 2\pi \approx 197\,\text{GHz}$ and its axial motion at $\omega_z / 2\pi \approx 50\,\text{MHz}$. The transitions are induced by driving the motional sideband on resonance ($\delta = 0$) using individual modes of a circular waveguide with varying radius $\rho_0$ and different mode indices $n$ and $\sigma$.}
	\label{fig:sideband_coupling_rate}
\end{figure}

By temporarily moving into the interaction picture, in which the free evolution defined by $\hat{H}_0$ is absorbed into the operators, and assuming the detuning is much smaller than the transition frequency ($\delta \ll \omega_+-\omega_z$), rapidly oscillating terms can be neglected in the typical rotating wave approximation. The interaction Hamiltonian for a single waveguide mode then reduces to
\begin{equation}
	\label{eq:theory:hamiltonian_sideband}
\hat{H}'_{s\sigma} = \hbar \frac{\Omega_{s\sigma}}{2} \left( e^{i\omega_{s\sigma} t} \, \hat{a}_z^\dagger \hat{a}_+ + e^{-i\omega_{s\sigma} t} \, \hat{a}_z \hat{a}_+^\dagger\right)\,,
\end{equation}
whose structure is often seen in the field of quantum optics and describes the periodic exchange of population between two modes, in this case the modified cyclotron motion and axial motion of an electron. The coupling strength is quantified by the Rabi frequency
\begin{subequations}
\label{eq:rabi_frequency_sideband}
\begin{align}
\Omega_{s1} &= a_{s1} \frac{q_e}{m_e} \sqrt{\frac{\hbar}{2 \epsilon_0 }} \frac{\omega_+ \chi_{s1}  c_{s1} }{ \rho_0 \sqrt{\omega_z (\omega_+ - \omega_-)}} \, \frac{k_{z,s1}^2}{\sqrt{\omega_{s1}}}\,,\\
\Omega_{s2} &= a_{s2} \frac{q_e}{m_e} \sqrt{\frac{\hbar}{2 \epsilon_0}} \frac{\omega_+ \chi_{s2}  c_{s2} }{ \rho_0 \sqrt{\omega_z (\omega_+ - \omega_-)}} \, k_{z,s2}\sqrt{\omega_{s2}}\,,
\label{eq:theory:rabi_frequency_sideband_TE}
\end{align}
\end{subequations}
which is plotted in \cref{fig:sideband_coupling_rate} for varying modes and waveguide radii, highlighting the need for careful mode selection and trap design.
The coupling leads to a periodic exchange between the average quantum numbers $\bar n_z (t)$ and $\bar n_+(t)$ described by the time evolution
\begin{equation}
\begin{split}
	\bar n_{+/z}(t) = \frac{\Omega^2}{\tilde{\Omega}^2} \sin^2\left(\frac{\tilde{\Omega}t}{2}\right) & \big[\bar n_{z/+}(0) - \bar n_{+/z}(0)\big]  \\
	&  + \bar n_{+/z}(0)\,,
\end{split}
\end{equation}
with the generalized Rabi frequency $\tilde{\Omega} = \sqrt{\Omega^2 + \delta^2}$ accounting for the effects of detuning.

In order to get an experimentally tangible expression for the coupling strength, the field amplitude $a_{s\sigma}$ is expressed in terms of power $P_{s\sigma}$ injected into that specific mode. Since the magnetic vector potential and mode functions are already defined to be normalized to single-photon energy in \cref{sec:theory:cyclotron_radiation}, the field amplitude scales as $a_{s\sigma} = \sqrt{n_{s\sigma}}$ where $n_{s\sigma}$ is the number of photons in a given mode.
Considering a section of waveguide with length $L$, the number of photons evaluates to $n = \Phi L / v_g$, with $\Phi = P / \hbar \omega$ being the incoming photon flux and $v_g = \text{d} \omega / \text{d} k = c^2 k_z / \omega$
being the group velocity. The field amplitude then becomes
\begin{equation}
a_{s\sigma} = \sqrt{\frac{L P_{s\sigma}}{\hbar c^2 k_{z, s\sigma}}}\,.
\end{equation}

When inserted into \cref{eq:rabi_frequency_sideband} to evaluate the sideband-coupling rate, the quantization length $L$ drops out due to the $\sqrt{V}$ term in the normalization constant $c_{s\sigma}$ defined in \cref{eq:normalization_constants}. The remaining dependence on the waveguide radius $\rho_0$,  as well as the behavior of different modes, is shown in \cref{fig:sideband_coupling_rate} and must be considered in the design of the trap geometry and millimeter-wave incoupling.

\subsection{Image charge coupling}
\label{sec:theory:image_charge_coupling}

\begin{figure}
	\includegraphics[width=\columnwidth]{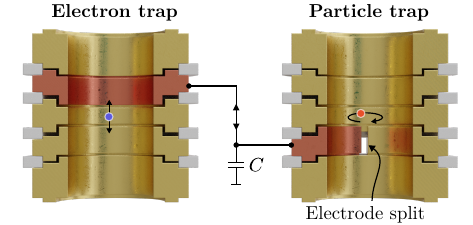}
	\caption{Schematic illustration of an electron and a particle-of-interest stored in two separate Penning traps being coupled by exchange of image charges between their pickup electrodes, shown in red. In the electron trap, an axially offset pickup electrode is employed to maximize sensitivity to the electron's axial motion, whereas in the particle trap, one half of a split electrode is used to enable coupling to the radial motion, as indicated by the motion-arrows. The capacitance $C$ is formed by the combined parasitic capacitance of the electrodes.}
	\label{fig:image_charge_coupling}
\end{figure}

A charged particle placed inside a Penning trap alters the local electrostatic field, leading to a redistribution of surface charges on the trap electrodes. Electrically connecting a specific electrode of the electron trap to a specific electrode of the particle trap, hereafter referred to as the ``pickup-electrodes'', allows surface charges to exchange and results in an effective coupling of the electron and particle-of-interest.
A schematic representation of this coupling is shown in \cref{fig:image_charge_coupling}.
To derive the corresponding interaction Hamiltonian, we start from the electrostatic interaction between the particles and the pickup-electrodes by applying Greens reciprocity theorem \cite{huSolvingBoundaryvalueElectrostatics2001} to two dual configurations of charge distributions within the coupled two-trap system:

In the first configuration, the electron and the particle-of-interest are placed within their respective traps at positions $\bm{r}_\e$ and $\bm{r}_\p$, respectively. The pickup-electrodes are assumed to carry zero charge and float at potential $V$ that is to be determined. The remaining electrodes are fixed to $0\,\text{V}$.
The second configuration considers the reverse scenario, in which the electron and the particle-of-interest are hypothetically assumed to have zero charge, while the pickup-electrodes and their connecting wire carry a combined net charge of $Q$. The remaining electrodes are again fixed to $0\,\text{V}$. In this configuration, the pickup-electrodes generate electrostatic potentials $\tilde{\Phi}_i (\bm{r}) = Q \phi_i(\bm{r}) / C$ within the electron and particle trap, respectively. Here, $C$ denotes the combined capacitance of the two connected pickup-electrodes and their connecting wire and $\phi_i (\bm{r})$ represent dimensionless weighting functions describing the spatial dependence of the potential as generated by the respective electrodes. For realistic electrode geometries, $\phi_i (\bm{r})$ typically lack easy analytical solutions and are therefore calculated numerically.

Combining the two cases using Greens reciprocity theorem yields
\begin{equation}
	V(\hat{\bm{r}}_\e, \hat{\bm{r}}_\p) = \frac{1}{C} \left[ q_\e \phi_\e(\hat{\bm{r}}_\e) + q_\p \phi_\p(\hat{\bm{r}}_\p) \right]\,,
\end{equation}
which represents the electrostatic potential at the combined pickup-electrodes created by the electron at $\bm{r}_\e$ and the particle-of-interest at $\bm{r}_\p$. This in turn creates a back-acting potential $\Phi_i = V(\hat{\bm{r}}_\e, \hat{\bm{r}}_\p) \phi_i(\hat{\bm{r}}_i)$ on the particles.

The Hamiltonian describing this electrostatic interaction is obtained by integrating the electrostatic energy density over all space
\begin{equation}
	\hat{H}' = \frac{1}{2} \int \rho(\bm{r}') \Phi(\bm{r}') \,\text{d}\bm{r}'\,,
\end{equation}
which reduces to an evaluation of the electrostatic potential at the position of the particles: 
\begin{equation}
	\hat{H}' = \frac{1}{2} V(\hat{\bm{r}}_\e, \hat{\bm{r}}_\p) \Big[ q_\e  \phi_\e(\hat{\bm{r}}_\e) + q_\p \phi_\p(\hat{\bm{r}}_\p) \Big]\,.
\end{equation}
Due to the small spatial extent of the particles' motions, the weighting functions $\phi_i(\bm{r})$ can be expressed as first-order Taylor expansions around their respective traps' center.
We limit the expansion to the Cartesian component exhibiting the steepest potential gradient, which corresponds to the direction to which an electrode is most sensitive to.
An axially offset electrode produces a large gradient in $z$ and is therefore used to couple to the axial oscillation in the electron trap, while a radially split electrode is used to couple to the radial oscillation components in either the $x$ or $y$ direction in the particle trap. 
In this expansion, the weighting potentials become 
\begin{equation}
	\phi_\e(\hat{\bm{r}}_\e) \approx  \hat{z}_\e/D_\e\,, \qquad \phi_\p(\hat{\bm{r}}_\p) \approx  \hat{x}_\p/D_\p\,,
\end{equation}
where the expansion coefficients $D_i$ are solely dependent on the trap geometries and are computed using finite-element methods. Typical values for $D_i$ are on the order of $1 \text{ to } 10\,\text{mm}$.
Expanding the interaction Hamiltonian yields
\begin{equation}
	\hat{H}' = \frac{1}{2C} \left[ \frac{q_\p^2 \hat{x}_\p^2}{D_\p^2} + \frac{q_\e^2 \hat{z}_\e^2}{D_\e^2} + \frac{q_\e q_\p}{D_\e D_\p} \left( \hat{x}_\p\hat{z}_\e + \hat{z}_\e\hat{x}_\p \right) \right]\,.
\end{equation}
The quadratic terms in $\hat{z}_e$ and $\hat{x}_p$ lead to small perturbations in the eigenfrequencies, which are neglected in the following.
Expressing the position coordinates as operators in their respective Fock basis
\begin{align*}
	\hat{z}_\e &= \sqrt{\frac{\hbar}{2 \omega_{z,\e} m_\e}} \left( \hat{a}_{z,\e}^\dagger + \hat{a}_{z,\e} \right)\,, \\
	\hat{x}_\p &= \sqrt{\frac{\hbar}{2 \omega_{1,\p} m_\p}} \left( \hat{a}_{+,\p}^\dagger + \hat{a}_{+,\p} + \hat{a}_{-+,\p}^\dagger + \hat{a}_{-,\p} \right)
\end{align*}
and applying a rotating wave approximation leads to the Hamiltonian
\begin{equation}
	\label{eq:theory:hamiltonian_image_charge_coupling}
	\hat{H}' = \hbar \frac{\Omega}{2} \left( \hat{a}_{z,\e}^\dagger \, \hat{a}_{+,\p} + \hat{a}_{z,\e} \, \hat{a}_{+,\p}^\dagger \right)\,.
\end{equation}
The coupling strength is quantified by the Rabi frequency
\begin{equation}
	\label{eq:rabi_frequency_ic_coupling}
	\Omega = \frac{q_\e q_\p}{C D_\e D_\p \sqrt{\omega_{1,\p} \, \omega_{z,\e}} \sqrt{m_\e m_p}}\,,
\end{equation}
which represents the frequency of periodic exchange between $\bar n_{z,\e}$ and $\bar n_{+,\p} $ at zero detuning $\delta = \omega_{z,\e} - \omega_{+,\p} = 0$. Plugging in representative numbers for coupling an electron to a $^{12}\text{C}^{6+}$ ion with $\omega_{1,\p} = \omega_{z,\e} = 54\,\text{MHz}$, $D_\e = D_\p = 2.7\,\text{mm}$ and $C = 2.5\,\text{pF}$ yields a Rabi frequency of $\Omega \approx 2\pi \cdot 30\,\text{mHz}$.
At non-zero detuning, the periodic exchange occurs at the modified Rabi frequency $\tilde{\Omega}^2 = \sqrt{\Omega^2 + \delta^2}$ and only exchanges a fraction $\Omega^2 / \tilde{\Omega}^2$ of the population.
In order to achieve sufficient coupling, the detuning $\delta$ must therefore be comparable to or smaller than the Rabi frequency. For the previous example, this would require a relative frequency stability of $\sim 10^{-10}$ for $\omega_{z,\e}$ and $\omega_{+,\p}$, which presents a significant challenge for experimental implementation.

This challenge is inherent to most sympathetic cooling schemes in Penning traps and has already been addressed in various ways \cite{tuTankCircuitAssistedCoupling2021,bohmanSympatheticCoolingTrapped2021,willSympatheticCoolingSchemes2022}.
One approach to enhance the coupling rate emerges from a partial cancellation of the parasitic capacitance $C$ by paralleling it with an inductor $L$ \cite{bohmanSympatheticCoolingTrapped2021}. While this technique allows for greatly enhanced coupling rates, it also gives rise to heating due to Johnson-Nyquist noise, which impairs the achievable steady-state temperature and is not considered for now.
Another approach to enhance the coupling rate involves increasing the number of coolant-particles involved in the coupling process, i.e. electrons in the present context. The common oscillatory mode of an ensemble of $N$ electrons behaves like a single particle with mass $N \cdot  m_\e$ and charge $N \cdot q_\e$, resulting in an enhanced Rabi frequency of $\Omega_N = \Omega\sqrt{N}$ when substituted into \cref{eq:rabi_frequency_ic_coupling}. Using $N=1000$ electrons increases the Rabi frequency to $\Omega_N \approx 1\,\text{Hz}$, which yields sufficient cooling performance as will be presented in \cref{sec:theory:full_coupling_chain}.

However, increasing the number of electrons gives rise to mutual Coulomb interactions, which results in a fluid-like (weakly correlated) non-neutral plasma with a rich internal mode structure \cite{dubinTheoryElectrostaticFluid1991, bollingerNonneutralIonPlasmas1994}.
The magnetic trapping field strongly magnetizes the plasma, resulting in an anisotropic velocity distribution between modes parallel and perpendicular to the magnetic field lines. Due to a many-electron adiabatic invariant \cite{oneilCollisionalDynamicsStrongly1985}, the equipartition rate between the parallel and perpendicular degrees of freedom (DOF) is vanishingly small. However, plasma modes can still couple within each DOF, which, in combination with imperfections in the trapping potential, could lead to heating of the center-of-mass modes.
While this effect is not expected to be significant at electron counts of $N = 1000$ \cite{weimerElectrostaticModesDiagnostic1994} and the rather efficient cooling provided by cyclotron radiation and sideband coupling would readily re-cool the center-of-mass modes as well as the coupled internal modes, it is still desirable to minimize any heating rate experienced by internal plasma modes.
Drag from asymmetries in the trapping potentials or collisions with neutral background gas causes heating and a radial expansion of the plasma \cite{dubinTrappedNonneutralPlasmas1999}, which might lead to shifts or broadening of the center-of-mass modes that might impair cooling performance. While expansion rates are expected to be small, especially at ultra high vacuum, high magnetic fields and low electron densities \cite{malmbergLongTimeContainmentPure1980, danielsonTorqueBalancedHighDensitySteady2005}, a rotating wall \cite{huangPreciseControlGlobal1998} or sideband drive \cite{weimerElectrostaticModesDiagnostic1994} might be employed to radially compress the plasma.

\subsection{Full cooling chain}
\label{sec:theory:full_coupling_chain}

Having established the individual Hamiltonians governing the dynamics of the distinct coupling stages, we now proceed to combine them into a single model. 
As the fully coupled system can no longer be solved analytically, numerical simulations are employed to investigate the system's dynamics.

The unitary time evolution of the full system is described by the sum of individual Hamiltonians described in the previous sections:
\begin{equation}
	\hat{H} = \hat{H}_{0,\e} + \hat{H}_{0,\p} + \underbrace{\hat{H}_\text{sb} + \hat{H}_\text{ic}}_{\hat{H}_1}\,.
\end{equation}
The free evolution of the electron and particle-of-interest are given by $\hat{H}_{0,\e}$ and $\hat{H}_{0,\p}$ as defined in \cref{eq:theory:hamiltonianCreationAnnihilation}. The terms $\hat{H}_\text{sb}$ and $\hat{H}_\text{ic}$ account for sideband-coupling using the Hamiltonian from \cref{eq:theory:hamiltonian_sideband} and image-current interaction using the Hamiltonian from \cref{eq:theory:hamiltonian_image_charge_coupling}, respectively. Combining these coupling terms into a single interaction Hamiltonian $\hat{H}_1$ and applying the unitary transformation $U = e^{i \hat{H}_{0,\e}t/\hbar} e^{i \hat{H}_{0,\p}t/\hbar}$ results in
\begin{equation}
	\label{eq:theoy:full_coupling_hamiltonian}
	\begin{split}
		\hat{H}_{1,\text{I}} = \hbar \frac{\Omega_\text{sb}}{2} & \left( e^{-i \delta_\text{sb} t}\hat{a}_{+,\e}^\dagger \, \hat{a}_{z,\e} + \text{H.c.} \right)
		\\
		+  \hbar \frac{\Omega_\text{ic}}{2} & \left( e^{-i \delta_\text{ic} t}\hat{a}_{z,\e}^\dagger \, \hat{a}_{+,\p} + \text{H.c.} \right)\,,
	\end{split}
\end{equation}
which describes the time-evolution of the system in the interaction picture. The complex exponentials account for a static detuning $\delta_\text{sb} = \omega_\text{sb} + \omega_{z,\e} - \omega_{+,\e}$ during sideband coupling , with $\omega_\text{sb}$ being the injected sideband-drive frequency, and for the detuning $\delta_\text{ic} = \omega_{z,\e} - \omega_{+,\p}$ during image-charge coupling.

The non-coherent time evolution arising from the interaction between the electrons' modified cyclotron motion and photons in the thermalized waveguide field is modeled using stochastic quantum jumps in the Monte Carlo wave function (MCWF) formalism \cite{plenioQuantumjumpApproachDissipative1998}. In this framework, the time evolution is split into a coherent integration of a non-hermitian Hamiltonian $\hat{H}_\text{MCWF} = \hat{H}_{1,\text{I}} - \frac{i\hbar}{2} \sum_i J_i^\dagger J_i$ and a stochastic application of collapse operators $J_i$ accounting for the non-coherent dynamics. By averaging over a large number of such stochastic quantum-trajectories, one reconstructs the time evolution of the system's density matrix.
In the present case, the emission/absorption of photons from/to the electron's modified cyclotron motion is modeled by the collapse operators
\begin{align*}
	J_\text{emi} &= \sqrt{\gamma \, \bar{n}_\text{th}} ~ \hat{a}_{+,\e}^\dagger\,, \\
	J_\text{abs} &= \sqrt{\gamma \left(\bar{n}_\text{th} + 1\right)} ~ \hat{a}_{+,\e}\,,
\end{align*}
where $\bar n_\text{th}$ represents the mean thermal photon population number in the field and $\gamma$ is the rate constant defined in \cref{eq:theory:rate_constants}.

For a realistic simulation of the system dynamics, fluctuations of the particles' eigenfrequencies caused by temporal voltage or magnetic field variations must be incorporated into the model.
While these fluctuations amount to at most $10^{-7}$ of the particles' eigenfrequencies and can therefore be safely neglected in the free particle Hamiltonian, they become relevant in the coupling dynamics. The fluctuations occur at time-scales of seconds, which allows them to be adiabatically approximated by introducing time-dependent detunings $\delta_\text{sb}(t)$ and $\delta_\text{ic}(t)$ in the interaction Hamiltonian \cref{eq:theoy:full_coupling_hamiltonian} by substituting
\begin{equation}
	\delta_i \, t \rightarrow \int_{-\infty}^t \delta_i (\tau) \, \text{d}\tau\,.
\end{equation}

In the numerical simulation, $\delta_\text{sb}(t)$ and $\delta_\text{ic}(t)$ are pre-computed for each quantum-trajectory by sampling from a normal distribution with variance $\sigma_\text{sb}^2$ and $\sigma_\text{ic}^2$ and band-limiting to $1\,\text{Hz}$. This stochastic approach integrates naturally into the MCWF formalism and converges in the limit of large numbers of simulated trajectories.

The numerical simulations are implemented using the QuTiP framework \cite{johanssonQuTiP2Python2013}.
The general simulation parameters are based on the theoretic framework derived in the previous sections and experimental boundary conditions, which will be described in detail in \cref{sec:experimental-setup}.
Assuming a magnetic field strength of $B_0 = 7\,\text{T}$ and a trap radius of $\rho_0 = 1.5\,\text{mm}$, the rate of cyclotron emission is calculated from \cref{eq:theory:rate_constants} to be $\gamma = 19\,\text{s}^{-1}$. According to the results from \cref{sec:theory:cyclotron_radiation}, the mean thermal photon population in the waveguide field is $\bar n_\text{th} = 0.4$. The sideband-coupling rate $\Omega_\text{sb}$ can be tuned experimentally over a broad range by adjusting the injected millimeter-wave power, and is thus treated as a variable parameter in the simulation.
The image-current coupling rate of $\Omega_\text{ic} = 2\pi \cdot 1\,\text{Hz}$ is adopted from the considerations in \cref{sec:theory:image_charge_coupling} for a $^{12}\text{C}^{6+}$ ion coupled to a cloud of 1000 electrons. The detuning-fluctuations are assumed to originate primarily from instabilities of the electron's axial frequency, which is why $\delta_\text{sb}(t)$ and $\delta_\text{ic}(t)$ are approximated to be coherent and sample the same noise vector with variance $\sigma_\text{sb}^2 = \sigma_\text{ic}^2 \eqqcolon \sigma^2$.

\begin{figure}
	\centering
	\includegraphics[width=\columnwidth]{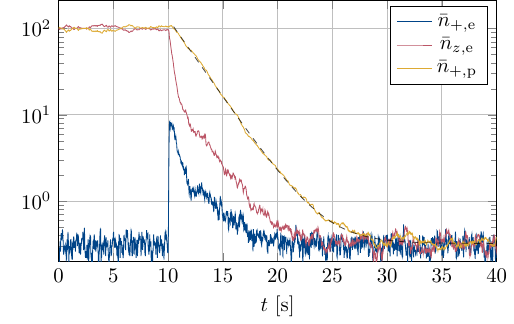}
	\caption{Numerical simulation to evaluate the average quantum numbers of the electron's modified cyclotron and axial motion ($\bar{n}_{+, \e}$ and $\bar{n}_{z, \e}$) and the particle-of-interest's modified cyclotron motion ($\bar{n}_{+, \p}$). The simulation uses a sideband coupling rate of $\Omega_\text{sb} = 2\pi \cdot 3\,\text{Hz}$ and detuning fluctuation with $\sigma = 1\,\text{Hz}$. Sideband coupling is initially disabled and switched on at $t=10\,\text{s}$. The dashed line represents an exponential fit to the modified cyclotron trajectory of the particle-of-interest, as used in the further analysis.}
	\label{fig:full_coupling_trajectory}
\end{figure}

An exemplary result from 100 averaged simulated trajectories is shown in \cref{fig:full_coupling_trajectory}, where the expectation values $\bar{n}_{+,\e}$, $\bar{n}_{z,\e}$ and $\bar{n}_{+,\p}$ are plotted over time.
The modified cyclotron motion of the carbon ion and the axial motion of the electron are initialized in a coherent state with $\alpha = \sqrt{100}$.
As can be seen, the expectation value of the ions modified cyclotron quantum number approximates an exponential decay, which is used to quantify the cooling speed using the time constant $\tau$ from an exponential fit $\bar{n}_{+, \p}(t) = e^{t/\tau} + \bar{n}_\text{th}$. Although this exponential behavior is not generally valid for all coupling parameters, it is a very good approximation within the probed parameter space.
The carbon ion's modified cyclotron motion thermalizes towards a thermal density matrix with $\bar{n}_{+, \p} = 0.4$, which corresponds to a temperature of $T_{+,\p} \approx 2\,\text{mK}$ when using $\omega_{+,\p} = 2\pi \cdot 54\,\text{MHz}$ from \cref{tab:eigenfrequencies}.
Generalizing to the range of particles listed in \cref{tab:eigenfrequencies}, the thermalized quantum number of 0.4 would translate to temperatures of $1.3 \text{ to } 4.0\,\text{mK}$ in the modified cyclotron motion.

\begin{figure}
	\centering
	\includegraphics[width=\columnwidth]{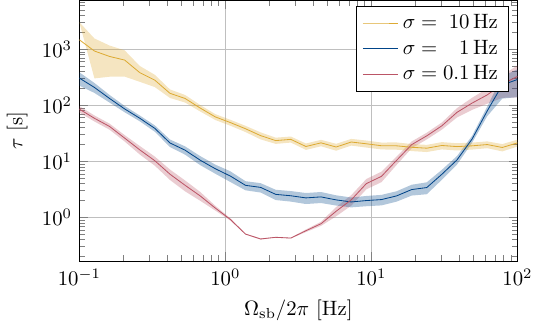}
	\caption{Simulated exponential decay constant $\tau$ of a $^{12}\text{C}^{6+}$ ion's average modified cyclotron quantum number for different detuning fluctuations $\sigma$ and varying sideband-coupling rate $\Omega_\text{sb}$. Each data point is obtained by fitting an exponential to an average of 100 simulated quantum trajectories as described in the text.}
	\label{fig:full_coupling_tau}
\end{figure}

\Cref{fig:full_coupling_tau} shows the exponential decay constant $\tau$ for different detuning variances $\sigma^2$ and sideband coupling strengths $\Omega_\text{sb}$.
For small detuning fluctuations $\sigma = 0.1\,\text{Hz}$, fastest cooling is achieved near $\Omega_\text{sb} \approx 2\pi \cdot 2\,\text{Hz}$, which corresponds to twice the image-charge coupling rate $\Omega_\text{ic}$ and coincides with the critical coupling condition between two damped harmonic oscillators. Increasing the sideband-coupling rate above this point degrades cooling performance due to over-damping of the electron axial motion.
For larger detuning fluctuations, the optimal operating point shifts towards greater sideband coupling strengths $\Omega_\text{sb}$. The increased damping of the electron axial motion broadens its linewidth, which is beneficial to accommodate for the larger swing of detuning during image-charge coupling.

\begin{figure*}
	\includegraphics[width=\textwidth]{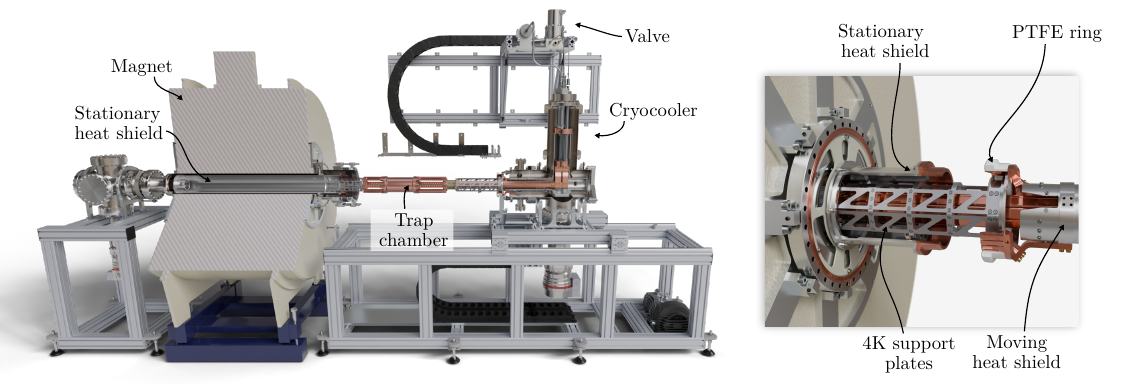}
	\caption{Rendering of the ELCOTRAP experiment in its ``pulled-out'' state. The inset shows the self-actuating mechanism that connects the cryocooler-side heat-shield to the stationary heat shield when cooled down. Thermal contact is established through contraction of the PTFE ring when cooling down.}
	\label{fig:elcotrap}
\end{figure*}

\section{The ELCOTRAP experiment}
\label{sec:experimental-setup}

The ELCOTRAP experiment is currently set up at the Max Planck Institute for Nuclear Physics in Heidelberg to test the proposed cooling technique and beyond that to serve as a general test platform for further technical developments in the field of Penning-trap physics. For this purpose, the experiment was specifically designed to allow easy and frequent access to the Penning trap and its accompanying electronics in order to enable quick development iterations.
An overview of the experiment is shown in \cref{fig:elcotrap}. A central component of the experiment is the superconducting magnet with a field strength of $7.05\,\text{T}$, whose horizontal bore allows the experiment to be laid out for easy user access.
The spatial homogeneity of the magnetic field measures $\Delta B / B = 1.2 \times 10^{-7}$ peak-to-peak within a cylindrical volume of $170\,\text{cm}^3$ in the central region of the magnet. The temporal stability is specified to $5 \times 10^{-8}\,\text{h}^{-1}$, which is surpassed in the long-term as shown by the results in \cref{sec:phase_1}. A pressure stabilization system for the magnet's nitrogen and helium reservoir was recently installed to further improve short-term stability, though it has not been fully commissioned yet. The magnet is connected to a helium recovery system to re-liquefy the evaporating helium.

The Penning-trap setup is cooled to near $4\,\text{K}$ using a closed-cycle cryocooler, which significantly simplifies cooldown and warmup procedures compared to alternative solutions employing liquid cryogens. The employed pulse-tube cryocooler provides two cooling stages with a cooling power of $36\,\text{W}$ at $48\,\text{K}$ at the first stage and $1.5\,\text{W}$ at $4\,\text{K}$ at the second stage. The thermally conductive connection to the Penning-trap chamber, which is placed in the region of highest magnetic field homogeneity right in the center of the magnet, is provided using bars made from oxygen-free high thermal conductivity (OFHC) copper and flexible copper braids to decouple misalignments arising from thermal expansion/contraction. The copper bars are arranged into a circular shape in order to provide a straight on-axis path for injecting millimeter waves as will be described in \cref{sec:phase_2}. The inner cryogenic setup is completely surrounded by a heat shield connected to the first stage of the cryocooler to reduce the heat-load from black body radiation. The heat shield is manufactured from aluminium to reduce weight and cost and is wrapped in 10 layers of super-insulation foil to further reduce the radiative heat load.
The cryogenic setup is placed into a vacuum chamber being pumped by two turbomolecular pumps achieving an ultimate pressure of $1 \times 10^{-8}\,\text{mbar}$ at room temperature and $< 1 \times 10^{-10}\,\text{mbar}$ during cryogenic operation.

A distinguishing feature of the ELCOTRAP experiment is the ability to slide the inner stage of the cryogenic setup out of the magnet, providing direct access to the trap chamber and cryogenic electronics. For this, the right part of the setup, including the cryocooler, a turbomolecular pump and the inner cryogenic setup as shown in \cref{fig:elcotrap}, is mounted on rails. The valve of the cryocooler is mounted on a separate rail-guided frame fixed to the wall of the laboratory in order to reduce vibrations. The heat shield is divided into a stationary tube that remains inside the magnet bore and a subassembly that is fixed to the cryocooler.
Both parts are automatically thermally connected when being cooled down using a self-actuating mechanism shown in \cref{fig:elcotrap}: A copper ring consisting of several flexible ``fingers'' is thermally connected to the cryocooler-side of the heat shield using flexible copper braids, which self centers coaxially onto a mating surface on the stationary heat shield tube when inserted into the magnet bore. A ring machined from PTFE is press-fit over the cryocooler-side ring, which applies an inward pressure onto the flexible fingers due to the high differential thermal contraction between PTFE and copper  \cite{ekinExperimentalTechniquesLowTemperature2006} when cooling down. This provides the necessary contact-force to thermally connect the stationary heat shield. The contact between the mating surfaces is further improved using a thin layer of \emph{Apiezon N} cryogenic grease.

The inner cryogenic setup is rigidly attached to the cryocooler-side ring using titanium plates specifically optimized for low thermal conductivity but high mechanical stiffness, providing a single fixed point of reference for the entire inner cryogenic setup. The ``left end'' of the inner cryogenic setup self-centers radially onto a registration cone on the heat shield, providing positioning repeatability while still allowing for thermal contraction in the axial direction.
When opening the experiment, the ``left end'' releases from the registration cone and continues to move along the heat shield on two small titanium wheels.
The process of sliding the cryogenic insert in or out of the magnet takes only 5 minutes and can be executed by one person alone. Despite its flexibility, the mechanical construction was designed for minimal deflection under self-weight which has been calculated to $< 200\,\text{\textmu m}$ at the trap chamber using finite element simulations.

The cooldown process of the setup can be started after an initial $2\,\text{h}$ pumpdown and achieves steady state temperatures of $6.4\,\text{K}$ and $100\,\text{K}$ at the trap chamber and heat shield after less than $24\,\text{h}$.
The high temperature of the heat shield compared to the cryocoolers specification of $\sim 48\,\text{K}$ is caused by a bottleneck in thermal conductivity at an aluminium tube connecting the first stage of the cryocooler to the heat shield. In the future, this part will be replaced by a copper tube. It is expected that the temperature of the inner cryogenic stage will also benefit from this upgrade.
The setup can be heated back to room-temperature via resistive heaters on both stages in less than 12 hours, resulting in a total cryo-cycle interval of approximately 1.5 days.

\subsection{Phase I: System commissioning}
\label{sec:phase_1}

\begin{figure}
	\includegraphics[width=\columnwidth]{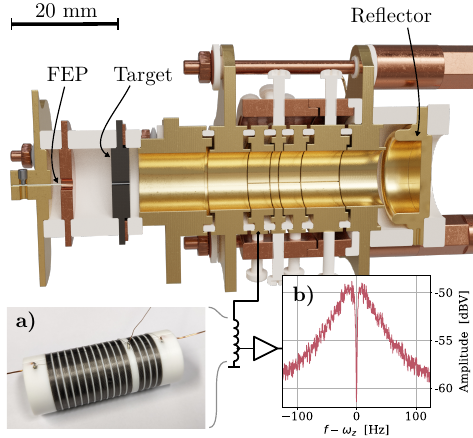}
	\caption{Rendering of the Phase-I trap design, consisting of a stack of gold-plated copper electrodes with an inner radius of $5\,\text{mm}$. On the left side, a field-emission point (FEP) inside an extraction electrode and a graphite target make up the primary part of the electron-beam ion source, which is completed by a reflector electrode on the right side. This allows carbon ions up to bare charge state to be produced directly in the Penning-trap potential, as is described in the text.
	One electrode is connected to a superconducting inductor shown in inset a), which forms the particle detection system together with a cryogenic amplifier. The resulting noise spectrum is shown in inset b), where the imprint of a single $^{12}\text{C}^{6+}$ ion is visible as a distinct dip at $\omega_z \approx 520\,\text{kHz}$.}
	\label{fig:phase_1}
\end{figure}

In the first experimental phase of the ELCOTRAP experiment, the general setup was commissioned and tested. During this phase, a simple Penning-trap electrode stack is used to test the general experiment infrastructure including the cryogenic systems, the particle detection system and the trap potential voltage source. The employed cylindrical Penning trap has a radius of $5\,\text{mm}$ and consists of 5 main electrodes that are designed to cancel electrostatic inhomogeneities up to 6th order. The trap design includes an integrated electron beam ion source (EBIS) based on the design of  \citet{alonsoMiniatureElectronbeamIon2006}, which allows to produce ions in situ. In the current configuration, the EBIS is equipped with a graphite target, allowing routine loading of carbon atoms up to its bare charge state. The EBIS also allows to load electrons, which can be trapped inside the same Penning trap when changing the polarity of the trapping voltages.

The trap assembly is mounted inside a chamber machined from OFHC copper which can be sealed using indium gaskets. A copper tube allows the chamber to be pumped down and ``pinched off'' prior to final assembly into the experiment, which provides the best attainable vacuum condition when cooling down due to cryopumping. Experiments employing this technique reported particle storage times of multiple years  \cite{sellnerImprovedLimitDirectly2017}. At a later stage, the pinch-off process was deliberately omitted, leaving the trap chamber open to the isolation vacuum through an orifice of $5\,\text{mm}$ diameter. Even though this is expected to worsen the ultimate vacuum inside the trapping region, only a single particle loss due to collisions with rest-gas atoms was witnessed during months worth of measurements, rendering this approach a worthwhile compromise between user-convenience and ultimate vacuum.

The confined particle is detected by observing its distinct imprint on the noise spectral density of a parallel RLC circuit connected between two trap electrodes \cite{winelandPrinciplesStoredIon1975}: The axial motion of the particle couples to the RLC circuit by means of image charges within the trap electrodes and effectively shorts the circuits Johnson Nyquist noise at a very narrow Bandwidth of typically a few Hertz around the axial frequency, leading to the dip-shape seen in \hyperref[fig:phase_1]{\ref{fig:phase_1}b}.
The signal-to-noise ratio (SNR) of this technique scales with the resistive contribution $R$, which stems solely from parasitic losses that must therefore be minimized. Since the capacitive contribution $C$ is fixed by the mutual and self-capacitance of the trap electrodes, the primary way to optimize the SNR is to minimize the losses within the inductor $L$. For phase~I of the ELCOTRAP experiment, a helical inductor wound using niobium-titanium (NbTi) wire on a hollowed out PTFE core mounted inside a NbTi housing is used, as shown in \hyperref[fig:phase_1]{\ref{fig:phase_1}a}. The thermal noise of the RLC circuit is amplified using a gallium-arsenide field-effect transistor amplifier optimized for cryogenic operation. 

Due to the narrow bandwidth of the RLC detection circuit, the axial frequency must be tuned to its resonance frequency by varying the depth of the trapping potential $V_0$, as indicated by \cref{eq:axial_frequency}.
The trapping voltage is provided by the ultra-stable voltage source StaReP  \cite{bohmUltrastableVoltageSource2016}, which shows relative voltage fluctuations lower than $10^{-7}$ over the relevant timespan of seconds to several minutes. Faster fluctuations and noise are minimized using RC filters placed at the $4\,\text{K}$ cryogenic stage with a time constant $\tau \approx 10\,\text{s}$.

\subsection{Phase II: Sideband coupling}
\label{sec:phase_2}

\begin{figure}
	\includegraphics[width=\columnwidth]{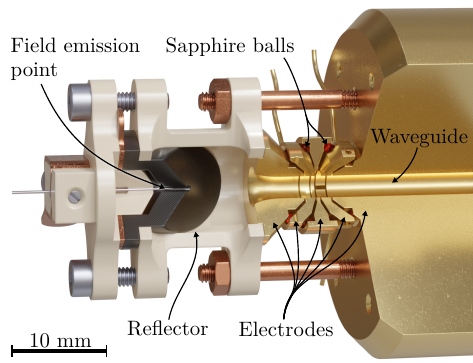}
	\caption{Sectional rendering of the Phase-II trap design. The trap itself consist of five gold-plated copper electrodes with inner radius of $\rho_0= 1.5\,\text{mm}$ spaced by $1\,\text{mm}$ diameter sapphire balls. The rightmost electrode acts as a structural component as well as a waveguide to deliver the millimeter waves to the trapping area. The millimeter waves exit the trap through an exponential horn in the leftmost electrode and are partially absorbed or scattered into the trap chamber using a lossy reflector.}
	\label{fig:phase_2}
\end{figure}

The second phase of the ELCOTRAP experiment is dedicated to implementing cyclotron-axial sideband coupling of electrons, which requires a special trap design, seen in \cref{fig:phase_2}.
As a primary requirement, this trap needs to enable millimeter-waves to be injected into the trap with as high field amplitude as possible. For this, the trap electrodes are designed to resemble a circular waveguide, which allows a traveling wave to be sent through it. As was described in \cref{sec:theory:sideband_coupling}, the only modes that provide the correct field pattern for sideband coupling of electrons located in the center of a waveguide are the $m=\pm 1$ modes. Therefore, it is necessary to maximize the injected power in one of these modes and prevent conversion into other ``useless'' modes. For this, the inner radius of the trap electrodes was minimized to $1.5\,\text{mm}$ under the constraints of manufacturability in order to limit the amount of available undesirable mode patterns and to achieve highest exchange rates, as shown \cref{fig:sideband_coupling_rate}. All upstream transitions were designed with smooth tapers in order to reduce the possibility of mode conversion.

The millimeter waves required for sideband coupling are produced outside the vacuum chamber starting with a signal generator producing a base band signal at $16.5\,\text{GHz}$, followed by a commercial frequency-multiplier to produce the required frequency of $197.5\,\text{GHz}$. The multiplied signal is subsequently amplified to $10\,\text{dBm}$ and can be attenuated using a connected programmable attenuator within a range of 0 to $70\,\text{dB}$.
When configured to drive the electron's motional sideband at maximum power, the millimeter-wave source produces phase noise with a power spectral density of $S_n = -103\,\text{dBm}/\text{Hz}$ at an offset of $\sim 50\,\text{MHz}$, which would excite the modified cyclotron frequency of the electrons to a temperature of $T = S_\text{N}/k_\text{B}$. Attenuators placed at room-temperature ($T_\text{room} \approx 290\,\text{K}$) and the inner cryogenic stage ($T_\text{cryo} \approx 4\,\text{K}$) are employed to reduce the effective field temperature, as seen by the modified cyclotron motion, to
\begin{equation}
	\begin{split}
		T_\text{eff} =  ~ & \eta_\text{cryo} \left[ \frac{S_\text{N}}{k_\text{B}} \, \eta_\text{room} + T_\text{room} \, (1-\eta_\text{room})\right]  \\
		& + T_\text{cryo} \, (1-\eta_\text{cryo})\,,
	\end{split}
\end{equation}
where $\eta_\text{room}$ and $\eta_\text{cryo}$ are the power transmission factors of the room-temperature and cryogenic attenuators, respectively, being related to their loss $L_\text{dB}$ via $\eta = 10^{-L_\text{dB}/10}$.
We initially aim for an effective field temperature of $T_\text{eff} = 8\,\text{K}$, which is realized using $57\,\text{dB}$ of room-temperature and $32\,\text{dB}$ of cryogenic attenuation.

The transmission of the millimeter waves to the trap is primarily achieved using hollow rectangular waveguides in standard size WR5, allowing only the fundamental $\text{TE}_{01}$ mode to propagate. However, the transition into the vacuum chamber requires the waveguide to be interrupted by a vacuum viewport, which is achieved using horn antennas with a gain of $25\,\text{dBi}$ on both sides of the viewport. This horn-to-horn transition also provides thermal isolation, which allows the vacuum side horn to be mounted directly to the $4\,\text{K}$ stage of the cryogenic setup. In order to prevent excessive radiative heating of the $4\,\text{K}$ stage, a PTFE lens thermally anchored to the heat shield is placed between the vacuum-side horn and the vacuum viewport. A benchtop measurement showed that this transition has a loss of $10\,\text{dB}$, which contributes to the room-temperature power transmission factor $\eta_\text{room}$.

A $1\,\text{m}$ long transmission line constructed from four straight WR5 waveguide sections connects the vacuum-side horn to the Penning trap. The transmission line was measured to introduce a loss of $12\,\text{dB}$, which is actually desired since it acts as part of the cryogenic attenuator to prevent phase-noise and room-temperature black-body photons from entering the trap. To reach the planned cryogenic attenuation of $32\,\text{dB}$, an additional fixed $20\,\text{dB}$ attenuator is placed into the transmission path.

Right before entering the trap chamber, a transition converts the rectangular WR5 waveguide profile into a circular profile matching the inner diameter of the Penning trap. This comparably large diameter is no longer single-moded and allows undesired modes to propagate. To prevent loosing power into these modes, a tapered transition that is sufficiently electrically long is used to minimize mode conversion.
The transition section passes through a cutout in the flange of the trap chamber and is screwed directly onto the long end-cap electrode of the Penning trap. While this direct connection opens the trap vacuum to the isolation vacuum through the hollow waveguide, it has the major benefit of sparing a second horn-horn transition into the trap-chamber. This design choice was made possible by the findings of ELCOTRAP phase I.

The calculated power received in the $\text{TE}_{11}$ mode in the center of the trap is $-79\,\text{dBm}$, or $13\,\text{pW}$. According to \cref{eq:theory:rabi_frequency_sideband_TE}, this corresponds to a sideband-coupling strength of $\Omega_\text{sb} \approx 2\pi \cdot 150\,\text{Hz}$, which is sufficient to reach the optimal coupling strength of $10 \text{ to } 100\,\text{Hz}$ observed in \cref{fig:full_coupling_tau}.

After the millimeter-waves have passed the trapping volume, they are radiated into free space using an exponentially horn that is machined into the last electrode. This horn gradually matches the wave impedance to that of free space, reducing reflections back into the trap and therefore prevents the formation of standing waves within the trapping region which could impair the sideband coupling rate.
The radiated waves are partially absorbed or scattered into the trap chamber using a reflector machined from carbon nanotube-filled PEEK material \cite{ElectricallyConductivePEEK} placed coaxially next to the last electrode. This reflector also acts as the extraction electrode for the field emission point that is mounted within it, which together form the electron source.
The trap electrodes are machined from OFHC copper and are plated with gold in order to prevent oxidation which could cause undesirable electrostatic patch potentials to form or might impair mm-wave propagation.
One of the electrodes is split radially in order to be able to apply a classical RF drive to sideband-cool the magnetron motion of the stored electrons.
The dimensions of the Penning trap electrodes themselves are optimized to provide the possibility to orthogonally tune the trap depth and trap anharmonicities \cite{gabrielseCylindricalPenningTraps1984}. Sapphire balls with a diameter of $1\,\text{mm}$ and a deviation from a perfect sphere smaller than $1\,\text{\textmu m}$ are used to separate the individual electrodes and the halves of the radially split electrode, offering a very high dimensional accuracy and predictable temperature behavior.
In comparison to sapphire rings, which are most commonly used in today's Penning-traps, the use of balls limits the amount of dielectric material between consecutive electrodes, which helps to minimize the electrode capacitance.

\begin{figure}
	\includegraphics[width=\columnwidth]{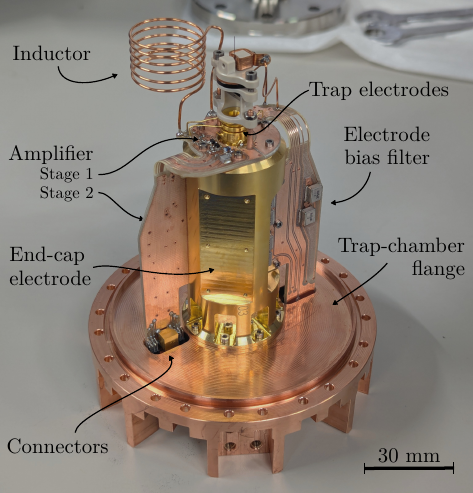}
	\caption{Photograph of the assembled Phase-II trap. The end-cap electrode is directly mounted onto the trap-chamber flange and enables the transmission of millimeter waves into the trapping region via its internal waveguide. Two semi-flex circuit boards integrate electrical connections to the electrodes, parts of the electrode DC bias filtering network and the cryogenic amplifier. The helical inductor, forming part of the detection system, is positioned close to the trap and amplifier to minimize parasitic effects.}
	\label{fig:phase_2_real}
\end{figure}

The electrons' axial motion is detected non-destructively using the dip technique described in \cref{sec:phase_1}.
A helical solenoid consisting of six turns of OFHC copper wire with a diameter of $1\,\text{mm}$, shown in \cref{fig:phase_2_real}, provides the inductance for the detection-system's RLC circuit. The low number of turns allows the coil to self-support, thereby minimizing the amount of lossy material in its vicinity.
A new cryogenic amplifier was designed under the strict requirement to use readily available commercial transistors, since all amplifier designs employed in cryogenic Penning-trap experiments known to us rely on devices that are now obsolete and difficult to source. After characterizing multiple transistors, the \emph{Skyworks SKY65050-372LF} high electron mobility transistor (HEMT) was selected due to its comparably low bias current requirements and stability characteristics at low temperatures. The amplifier features a cascode input stage to minimize the parasitic capacitance and feedback to the input, followed by a second stage in common-source topology to further increase the total gain. Impedance matching between the two amplifier stages and matching of the output transistor to $50\,\Omega$ line impedance are realized using reactive networks, which also provide transistor biasing. The matching networks are tuned to achieve unconditional stability across the whole operating temperature range. Measurements at cryogenic temperatures showed a gain of $34\,\text{dB}$ and an input-referred noise density of $0.48\,\text{nV}/\sqrt{\text{Hz}}$ at a frequency of $54\,\text{MHz}$, while consuming $< 2\,\text{mW}$ of power.
The amplifier is integrated onto a semi-flex circuit board, shown in \cref{fig:phase_2_real}, which allows the first stage to be placed in close proximity to the trap electrodes to minimize parasitics. Only dedicated non-magnetic components were used to preserve magnetic field homogeneity. The semi-flex circuit boards also provide electrical wiring of the trap electrodes, filtering of the electrode bias supply lines and connections to another circuit board located on the opposite side of the trap-chamber flange through cutouts.

\subsection{Phase III: Image-charge coupling}

In the third experimental phase, the image-charge-mediated coupling between the electrons and the particle-of-interest will be implemented and combined with the sideband-coupling as tested in phase II. This completes the full coupling chain and represents the final development step of the ELCOTRAP experiment.
The design of the trap-setup is still in progress and will be guided by the results and optimizations found during phase II.
It will feature two separate traps for storing the electrons and the particle-of-interest. The design will be highly optimized to enhance the image-charge-mediated coupling between the two traps, which is primary achieved by minimizing the combined electrode capacitance $C$ and the effective electrode distances $D_\e$ and $D_\p$, as indicated by \cref{eq:rabi_frequency_ic_coupling}.
The combined electrode capacitance is minimized by careful optimization of the electrode geometries and using individual sapphire balls for electrode separation, as used in Phase II.
The effective electrode distances are primarily minimized by reducing the overall trap-dimensions. Furthermore, multiple electrodes may be coupled into a combined virtual pickup electrode to further lower the effective electrode distance.
FEM simulations of preliminary conceptual designs indicate that a combined trap capacitance of $C \approx 2.5\,\text{pF}$ and effective electrode distances of $D \approx 2.7\,\text{mm}$ for both the electron and particle-trap are realizable. These values were used to calculate the image-charge coupling rates in \cref{sec:theory:image_charge_coupling}.

\begin{figure}
	\includegraphics[width=\columnwidth]{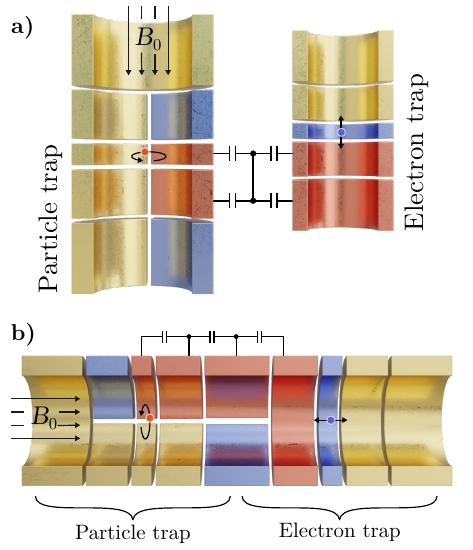}
	\caption{Different topologies for image-charge-mediated coupling between two traps. Figure a) shows a concept where the electron trap is placed laterally offset to the particle trap in the same magnetic field. Multiple pickup electrodes (red) are capacitively coupled to lower the effective electrode distance per trap. The blue electrodes are specifically decoupled from the biasing circuit, leaving them essentially floating at the coupling frequencies, which helps reducing the parasitic electrode capacitance.
	Figure b) shows a concept where both traps are aligned coaxially and share one of their end-cap electrodes. A combination of full- and split electrodes are capacitively connected to allow effective coupling between the electron's axial motions and the particle-of-interest's modified cyclotron motion.}
	\label{fig:phase_3}
\end{figure}

The electrical connection required to mediate image-charges between the two traps is implemented capacitively. This enables coupling at frequencies of $10 \text{ to } 100\,\text{MHz}$, while simultaneously allowing DC-biasing of the electrodes.
Two different trap-arrangements are considered, both shown in \cref{fig:phase_3}. In the first design concept, the electron trap is positioned laterally offset from the particle trap, enabling a more compact layout and simplifying integration with additional measurement traps. The magnet’s homogeneous field region is sufficiently large to accommodate the offset trap without compromising field uniformity. The offset electron trap decouples electron loading and millimeter-wave injection from the particle trap and any potentially coaxially mounted measurement traps, reducing interference and simplifying experimental configurations.
The second concept follows a coaxial alignment of both traps, in which particle-of-interest and electrons are stored in different regions of a common electrode stack and are coupled using capacitively connected electrode segments.
This approach integrates seamlessly with stacked traps of conventional Penning-trap experiments and has minimal radial space requirements.
The two different configurations support integration into a variety of Penning-trap architectures to optimally accommodate diverse experimental objectives and spatial constraints.

\section{Conclusion and Outlook}

This work introduces a new approach for achieving ultra-low temperatures of arbitrary charged particles confined within Penning traps.
A theoretical foundation is developed based on a quantum-mechanical description of the system, which encompasses the radiation of the electron's modified cyclotron motion into specific electromagnetic field modes of a cylindrical Penning trap as well as the sideband- and image-current mediated coupling mechanisms between different motional modes.
A numerical simulation of the particle-of-interest's time evolution yield\textit{s} a cooling time constant of $20 \text{ to } 200\,\text{s}$, depending on the experimental parameters. The cooling time is primarily limited by the comparably weak image-charge interaction and fluctuations in the Penning trap's bias voltage supply.
The final temperature in the modified cyclotron motion of a broad range of highly charged ions or (anti)protons was calculated to range from $1.3 \text{ to } 4.0\,\text{mK}$. In contrast to the most commonly used resistive cooling technique, which typically yields temperatures on the order of $1 \text{ to } 500\,\text{K}$ in the modified cyclotron motion, this approach represents a promising opportunity for the field of Penning-trap experiments.
While recent developments towards sympathetic cooling using clouds of laser-cooled beryllium ions pursue the same goal of achieving ultra-low temperatures in Penning traps \cite{bohmanSympatheticCoolingTrapped2021, tuTankCircuitAssistedCoupling2021}, the simplicity of using electrons instead of beryllium ions and millimeter-waves instead of cooling-lasers offers significant experimental advantages.
Continued efforts on these complementary approaches will contribute to address the shared challenges inherent to image-charge-mediated sympathetic cooling techniques and pave the way towards the millikelvin regime in Penning traps.

The first implementation of the proposed technique is currently being carried out in successive phases at the dedicated ELCOTRAP experiment. 
Phase I successfully demonstrated the initial systems operation including cryogenic cycling, a custom ultra-stable voltage source, in-trap ion loading and single-particle detection.
Phase II focuses on the implementation of sideband-coupling of the electron's motional degrees of freedom using millimeter-waves at $197\,\text{GHz}$. A highly specialized trap design and millimeter-wave transmission path are currently being commissioned and first results are expected in the near future.
Phase III will integrate the complete cooling chain by establishing image-charge coupling between two spatially separated traps.
This final step allows to transfer energy from the particle-of-interest to the cloud of well-cooled electrons and ultimately brings it to the targeted ultra-low temperatures.

\section*{Acknowledgments}

We thank Sven Sturm for valuable input and initial ideas that inspired this project. We also acknowledge his independent research on sideband cooling of positrons as part of the \textsc{Lsym} experiment \cite{holzenkampDesignMicrowaveCavity2025,raabDesigningCharacterizationNarrow2024}, which provided a basis for fruitful discussions.
This work is part of and funded by the Max Planck Society.
Furthermore, this project has received funding from the
German Research Foundation (DFG) Project-ID 273811115—SFB
1225 ISOQUANT. This work comprises parts of the Ph.D. thesis work
of J. Herkenhoff to be submitted to Heidelberg University, Germany.

\section*{Data availability}
The data that support the findings of this article are openly available \cite{herkenhoffDataSympatheticCooling2025}.

\end{document}